\documentclass[sigchi]{acmart}

\usepackage{booktabs} 

\usepackage{balance}

\AtBeginDocument{%
  \providecommand\BibTeX{{%
    \normalfont B\kern-0.5em{\scshape i\kern-0.25em b}\kern-0.8em\TeX}}}

\setcopyright{acmlicensed}
\copyrightyear{2019}
\acmYear{2019}
\acmConference[ICMI '19]{2019 International Conference on Multimodal Interaction}{October 14-18, 2019}{Suzhou, China}
\acmBooktitle{2019 International Conference on Multimodal Interaction (ICMI '19), October 14-18, 2019, Suzhou, China}
\acmPrice{}
\acmDOI{10.1145/3340555.3353744}
\acmISBN{978-1-4503-6860-5/19/05}
\begin{document}

%
\title[An Open Embodied Avatar]{A High-Fidelity Open Embodied Avatar with Lip Syncing and Expression Capabilities}

\author{Deepali Aneja}
\email{deepalia@cs.washington.edu}
\affiliation{%
  \institution{University of Washington}
  \streetaddress{}
  \city{Seattle}
  \state{Washington}
  \postcode{}
}

\author{Daniel McDuff}
\email{damcduff@microsoft.com}
\affiliation{%
  \institution{Microsoft Research}
  \streetaddress{}
  \city{Redmond}
  \state{Washington}
  \postcode{}
}

\author{Shital Shah}
\email{shitals@microsoft.com}
\affiliation{%
  \institution{Microsoft Research}
  \streetaddress{}
  \city{Redmond}
  \state{Washington}
  \postcode{}
}

%
\renewcommand{\shortauthors}{Aneja et al.}

%
\begin{abstract}
Embodied avatars as virtual agents have many applications and provide benefits over disembodied agents, allowing nonverbal social and interactional cues to be leveraged, in a similar manner to how humans interact with each other. We present an open embodied avatar built upon the Unreal Engine that can be controlled via a simple python programming interface. The avatar has lip syncing (phoneme control), head gesture and facial expression (using either facial action units or cardinal emotion categories) capabilities. We release code and models to illustrate how the avatar can be controlled like a puppet or used to create a simple conversational agent using public application programming interfaces (APIs). \newline
GITHUB link: \url{https://github.com/danmcduff/AvatarSim}
\end{abstract}

%
%


%

\maketitle

\section{Introduction}
Virtual agents have many applications, from assistance to companionship. With advancements in machine intelligence, computer graphics and human-computer interaction, agents are becoming more and more capable of real-time interaction with the user, synthesizing speech and expressions, style matching in conversations, tracking and completing tasks as assistants. How we represent an intelligent system via a user interface has implications for how the user perceives the system and associates or attributes intelligence with it. One line of research emphasizes that embodied agents offer several advantages over non-embodied dialogue systems. An agent that has a physical presence means that the user can look at it. Cassell~\cite{cassell2001embodied} argues that embodiment can help locate intelligence for users and reconcile both interactional and propositional intelligence. Embodied agents have greater expressive capabilities and naturally embodied agents have been used to explore the role of social signals~\cite{cassell2000embodied} and emotions~\cite{pelachaud2002subtleties} in human-computer interaction.

\begin{figure}[t]
  \centering
  \includegraphics[width=0.82\linewidth,keepaspectratio]{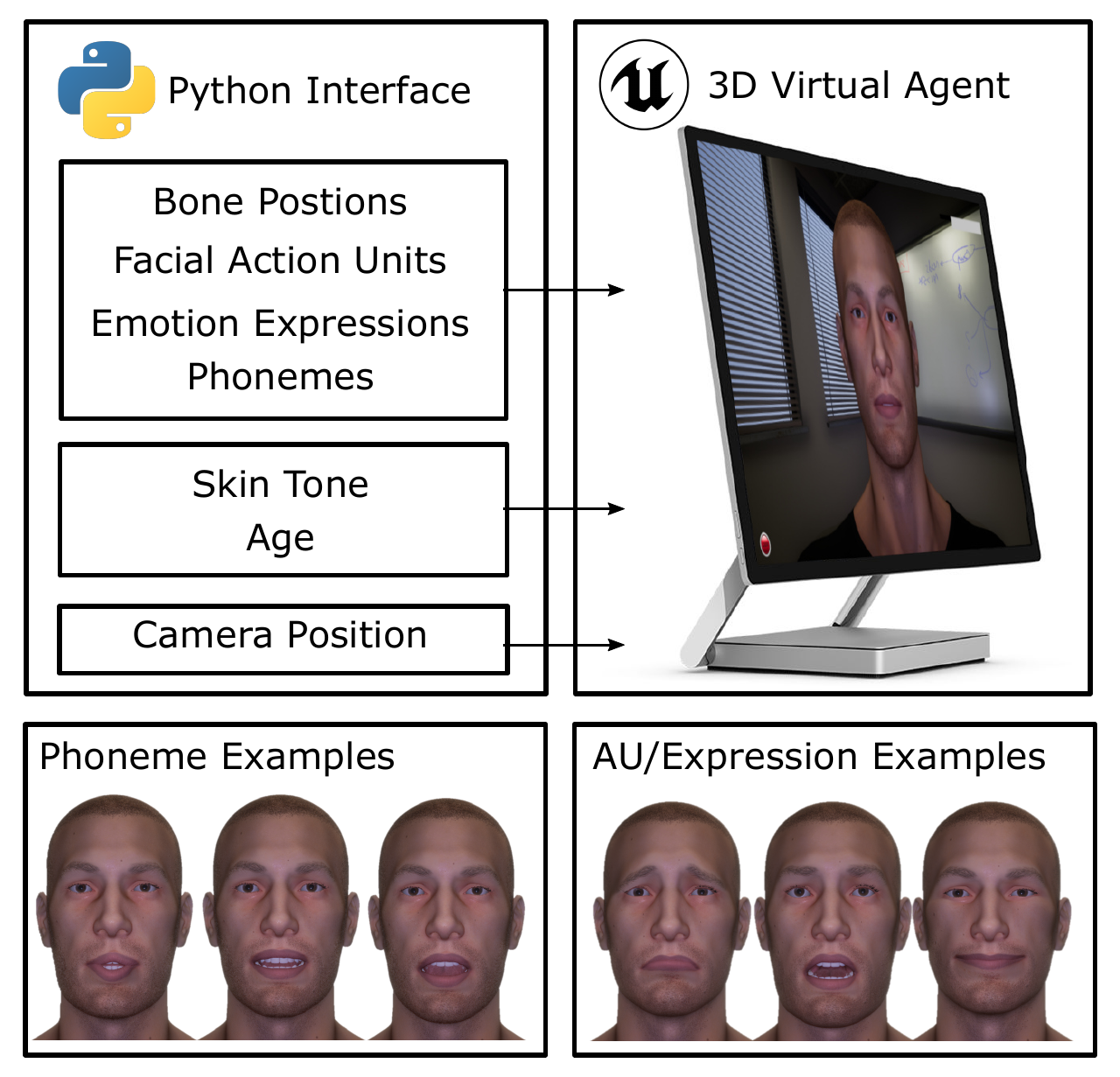}
  \caption{We present an open embodied avatar with lip syncing and expression capabilities that can be controlled via simple python interface. We provide examples of how to combine this with publicly available speech and dialogue APIs to construct a conversational embodied agent.}
  \label{fig:teaser}
  \vspace{-0.3cm}
\end{figure}

Improvements in speech recognition~\cite{xiong2018microsoft}, dialogue generation~\cite{serban2016building,huber2018emotional}, emotional speech synthesis~\cite{skerry2018towards,ma2018generative} and computer graphics have made it possible to design more expressive and realistic conversational agents. However, there are still many uncertainties in how best to design embodied agents. Building a state-of-the-art system for performing research to address these questions is non-trivial.

\begin{figure}[t]
  \centering
  \includegraphics[width=\linewidth]{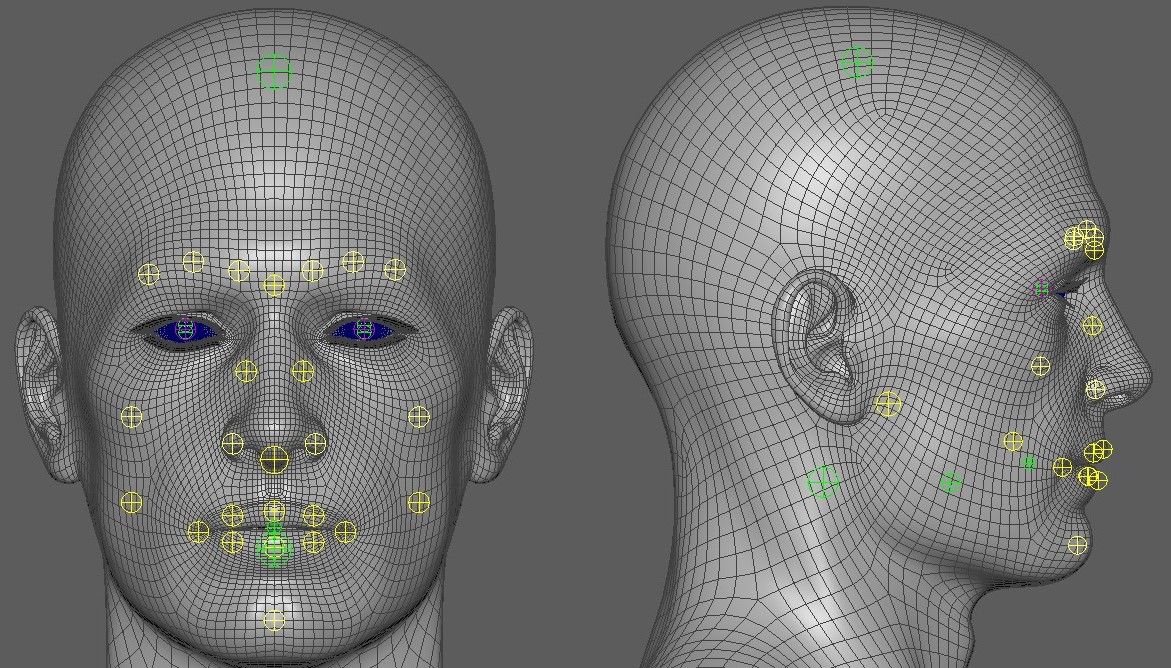}
  \caption{The locations of controllable bone positions on the avatar. Each position can be controlled with 6 degrees of freedom. See Table~\ref{tab:freq} for descriptions.}
  \label{fig:avatar}
  \vspace{-0.3cm}

\end{figure}

Open code, data, and models democratize research, helping advance the state-of-the-art by enabling a researcher to build upon existing work. LUCIA~\cite{interuniversitariolucia} is an MPEG-4 facial animation system that can copy a real human being by reproducing the movements of passive markers positioned on his face. Several architectures have been proposed for designing and building embodied conversational agents (ECA). AVATAR~\cite{santos2011avatar} is an architecture to develop ECAs based on open source tools and libraries. FACSvatar~\cite{van2018facsvatar} processes and animates action units in real-time. The framework is designed to be modular making it easy to integrate into other systems and has low latency allowing real-time animations. FACSHuman~\cite{gilbert2018facshuman} is a similar tool that allows animation of a 3-dimensional avatar using the facial action coding system (FACS)\cite{Ekman1972}. The system, however, does not allow for easily constructing real-time multimodal ECAs.  

These systems have been very valuable. However, there are some limitations. First, driving the non-verbal behavior of the agent is often not simple. For example, LUCIA~\cite{interuniversitariolucia} requires the user to wear reflective markers. 
In \cite{mcdonnell2012render}, people rated the realistic characters or very cartoon-like characters as more appealing or pleasant, but not characters which lie in the middle of 'abstract to realistic' scale. GRETA is an autonomous and expressive agent that is somewhat realistic in appearance and has numerous capabilities for social and emotional expression in addition to dialogue~\cite{poggi2005greta}. We present a freely available high-fidelity embodied avatar that has lip syncing, head gesture and facial expression capabilities. It can be controlled via a simple python interface, using either bone positions, action units or basic expressions. The goal is to make it easier for people to develop ECAs.

The contributions of this paper are: (1) to present a high-fidelity open embodied avatar that is capable of lip syncing and facial expression control via a simple python interface, using either bone positions, FACS or basic expressions, (2) to provide an example of an end-to-end conversational agent interface via publicly available APIs, and (3) to show how expression transfer can be used to control the avatar.

\section{The Avatar}
We created the avatar within the AirSim \cite{shah2018airsim} environment using Unreal Engine 4 (UE4)~\cite{games2007unreal}. The avatar, a head, and torso is placed in a room with multiple light sources (windows, overhead lights, and spotlights). Pictures of the avatar are shown in Figure~\ref{fig:teaser}. The avatar's appearance can be manipulated in several ways. The position and orientation of 38 ``bone" positions, which are facial locations (or landmarks) can be controlled, each with six degrees of freedom. We also created 24 FACS action unit presets \cite{Ekman1972} and 19 phonetic mouth shape presets (visemes) for the avatar. Each of these variables can be controlled in real-time using a simple Python interface. Once the avatar executable is running and a client connection formed between the python instance the positions of bones, the action unit and phoneme presets or the location of the camera can be updated as frequently as needed. The avatar controls are described as follows: \\

\textbf{Bone Position Controls.}
The position (x$_b$,y$_b$,z$_b$) and orientation ($\sigma_b1$,$\sigma_b2$,$\sigma_b3$) of each of the 38 bone positions can be set. Figure~\ref{fig:avatar} shows the location of the bone positions and Table~\ref{tab:freq} lists their names. The green points indicate joints that control rigid head and jaw motions. The yellow points indicate landmarks that control specific facial regions (e.g., eyebrows). The bones positions can be referred to via these names (e.g., LUpperEyelid for the left upper eyelid) or by their integer as specified in our documentation. 
\\

\textbf{Facial Action Controls.}
FACS is the mostly widely used and objective taxonomy for describing facial behavior. Therefore, in addition to control of bone positions the avatar has controls for 24 facial action unit presets (see Table~\ref{tab:freq}). 
\\

\begin{table}[t]
  \caption{The avatar has bone position, facial action unit and phoneme controls. M=middle, L=left, R=right, U=upper, L=lower, C=corner, I=inner, O=outer}
  \vspace{-0.4cm}
  \label{tab:freq}
  
  \begin{tabular}{llc}
    \toprule
    Bone Positions & Action Units & Phonemes \\
    \midrule
    U. Cheek (L \& R) & 1: I. Brow Raiser  & ae \\
    Mid U. Lip (M, L \& R) & 2: O. Brow Raiser  & b \\
    Mid L. Lip (M, L \& R) & 4: Brow Lowerer & d \\
    Lip Corner (L \& R) & 5: U. Lid Raiser & ee \\
    U. Nose (M, L \& R) & 6: Cheek Raiser & ngnk \\
    O. Brow (L \& R) & 7: Lid Tightener & eh \\
    Mid Brow (L \& R) & 9: Nose Wrinkler & ar\\
    I. Brow (L \& R) & 10: U. Lip Raiser & f \\
    Nostril (L \& R) & 11: Nasolabial Deep. & g \\
    Cheek Dimple (L \& R) & 12: Lip C. Puller & h \\
    Eye (L \& R) & 13: Sharp Lip Puller & i \\
    U. Eyelid (L \& R) & 14: Dimpler & k \\
    L. Eyelid (L \& R) & 15: Lip C. Depressor & l \\
    Jaw & 16: L. Lip Depressor & m \\
    Chin & 17: Chin Raiser & n \\
    Tongue Base & 18: Lip Pucker & oh \\
    Tongue Tip & 20: Lip Stretcher & ooo \\
    Chest  & 22: Lip Funneler & oh \\
    Neck  & 23: Lip Tightener & m \\
    Head  & 24: Lip Pressor &  \\
    Root  & 25: Lips Part &  \\
    Hair  & 26: Jaw Drop &  \\
      & 27: Mouth Stretch &  \\
      & 28: Lip Suck &  \\
  \bottomrule
\end{tabular}
\vspace{-0.4cm}
\end{table}

\textbf{Expression Controls.}
A set of basic emotional expressions can be created via combinations of action units (e.g., using EMFACS~\cite{friesen1983emfacs}) or using expression retargeting (Section 3). The avatar is also capable of blinking the eyes by controlling the upper and lower eyelid bone position controls.
\\

\textbf{Lip Syncing Controls.}
For conversational applications lip syncing is enabled via phoneme control. In our example scripts, we show how a phoneme classifier~\cite{huggins2006pocketsphinx} can be first used to identify phonemes from an audio segment and then used to control the lips in synchrony with the audio playback. Nineteen phoneme presets are included, while there are 44 phonemes in English the 19 presets can be used quite effectively for most dialogue. 
\\

\textbf{Head Gesture and Appearance Controls.}
The head rotations can be varied continuously by controlling the yaw, pitch and roll between -1.0f to 1.0f. The skin tone and age of the avatar can be controlled via independent parameters.  The skin tone varied from light (0) to dark (1), and the skin age texture can be varied from youthful (0) to old (1).
\\

\begin{figure*}[t]
  \centering
  \includegraphics[width=\textwidth]{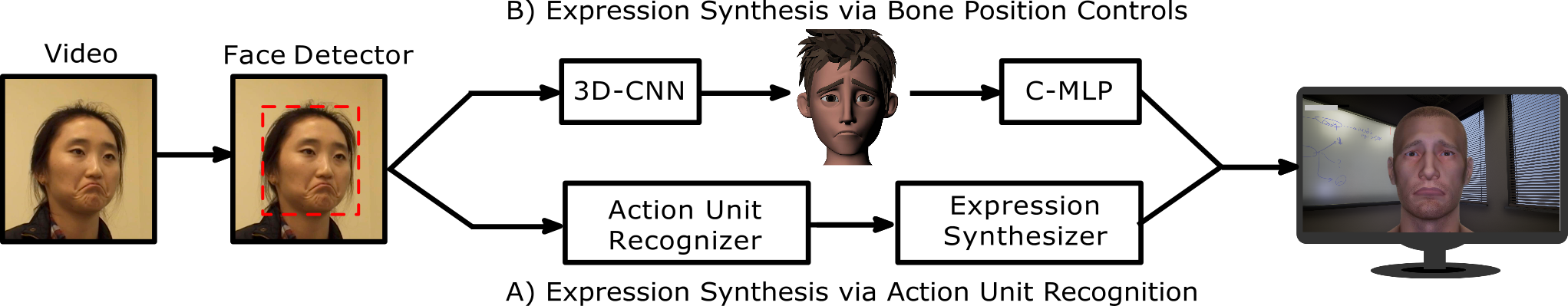}
  \caption{We present two pipelines for retargeting an expression from a human to the avatar. A) Expression Synthesis via Action Units Controls. Our pipeline takes a human video as an input, recognizes 12 FACS Action Units from the user's detected face and synthesizes the same expression on the avatar's face. B) Expression Synthesis via Bone Position Controls. Our pipeline takes a human video as an input, detects the user's face, generates 3D parameters on the primary character 'Ray' through a 3D-CNN and synthesizes the same expression on the avatar's face. See Section 3.2 for details for the Action Unit Recognizer, Expression Synthesizer, 3D-CNN and Character Multi-layer Perceptron (C-MLP).}
  \label{fig:au_pipeline}
\end{figure*}

\textbf{Camera Controls.}
The camera position within the virtual environment that governs the perspective from which the agent will be captured can be controlled with six degrees of freedom (position (x$_c$,y$_c$,z$_c$) and orientation ($\sigma_c1$,$\sigma_c2$,$\sigma_c3$)).  

\section{Applications}
\subsection{Conversational Agent}
To illustrate how the avatar can be used as a conversational agent, we provide an end-to-end pipeline for conversational dialogue, similar to~\cite{hoegen2019end}. The pipeline uses public APIs for natural language generation and speech synthesis from Microsoft.  We created a demonstration script that calls a natural language API to generate utterances and then uses a text-to-speech service to return a synthesized voice. We convert this voice into a sequence of visual groups of phonemes (visemes) using PocketSphinx~\cite{huggins2006pocketsphinx}. Finally, we synchronously play the audio and drive the avatar using the phoneme presets to achieve a simple lip syncing example of a conversational agent. Python scripts for running this conversational agent are provided with the agent software.  

\subsection{Facial Expression Retargeting}
In virtual embodied agent interactions, whether for conversation or not, expressions and head motions are important for creating a more natural and lifelike agent. Our next example shows how expressive faces can be generated. We provide two interfaces for driving the expressions of the avatar (see Figure~\ref{fig:au_pipeline}), one using bone controls and the other using the FACS presets. In both cases, we utilize existing deep learning based frameworks to perform the expression transfer i.e., synthesizing the avatar's face with the expressions of a human subject in the input video. This creates the effect of facial expression mimicry. We provide scripts and a FACS-based trained model for systematically generating faces. The facial expressions on the avatar can be synthesized as follows: \\

\textbf{Expression Synthesis via Facial Action Units.}
We describe an end-to-end pipeline to retarget a human's facial expression to the avatar's face using facial action unit controls as shown in Figure \ref{fig:au_pipeline} (A). The pipeline utilizes a webcam attached to the computer and processes the video frames. First, we detect the face in the input video. We use an open source HoG-based face detector \cite{dlib09}. Next, we analyze the facial region of interest (ROI), defined by the face detection bounding box, using the Facial Action Unit Recognizer. The output from the Action Unit Recognizer is used to synthesize the expressions on the avatar's face by the Expression Synthesizer.

\subsubsection{Facial Action Unit Recognizer}
We trained a Convolutional Neural Network (CNN) \cite{he2016deep} on a large facial expression dataset \cite{fabian2016emotionet} for recognizing 12 facial action units - 1, 2, 4, 5, 6, 9, 12, 17, 20, 25, 26 (refer to Table~\ref{tab:freq} for descriptions) and 43 (Eyes Closed). The Action Unit Recognizer returns a 12-dimensional feature vector representing the probability of each action unit. The model was trained using PyTorch \cite{paszke2017automatic} to minimize the binary cross-entropy loss with an average F1-score of 0.78. More details about the training and evaluation of the model can be found in~\cite{auModel}. Alternatively, the avatar could be used with an off-the-shelf FACS AU detection SDK, such as~\cite{mcduff2016affdex}.

\subsubsection{Expression Synthesizer}
We use the facial action unit probability scores from the Facial Action Unit Recognizer to control the avatar's facial action controls and mimic the human facial expressions. The raw probability scores can be used directly to drive the actions, or these values can be smoothed or rounded first. 
Python scripts for running the expression synthesis via facial action units are provided with the agent software along with the trained Action Unit Recognizer model. 
\\

\textbf{Expression Synthesis via Bone Position Controls.}
We describe an end-to-end pipeline to retarget a human's facial expression to the avatar's face using the bone position controls as shown in Figure \ref{fig:au_pipeline} (B). As with the previous approach face detection is first performed on the input video and the bounding box used to define the ROI. We then use a multi-stage deep learning system ExprGen \cite{aneja2018learning} to retarget facial expressions from the human to the avatar. The pipeline takes 2D images of human facial expressions as input and generates 3D parameters of the avatar. We have two separate components in our pipeline: Human to Primary Character expression transfer and Primary Character to Avatar expression transfer. 

\subsubsection{Human to Primary Character transfer} The 3D-CNN produces parameters for a human to a primary  character expression transfer in 3D including both perceptual and geometric similarity as shown on 'Ray' character in Figure \ref{fig:au_pipeline} (B) (middle).

\subsubsection{Primary Character to Avatar transfer} We trained a Character Multi-layer Perceptron (C-MLP) to transfer the expression of the primary character to the avatar. The C-MLP was trained to minimize the mean square error loss between the primary character parameters and the avatar parameters. More details about training the model can be found in \cite{aneja2016modeling, aneja2018learning}.

Some of the limitations of our work are: The intensity of FACS action units is not validated, and the synthesis of composite facial expressions using multiple facial action units along with phoneme based lip syncing is currently rule-based and not validated for expression clarity. When using the avatar as a conversational agent along with expression retargeting, it can result in implausible facial movement due to the conflict between lip syncing controls and expression controls. To avoid such a scenario, we control the expressiveness of the avatar's upper face only (above the mouth) while lip syncing. However, the avatar allows for exploration of more advanced solutions to this problem.  In future research, we can use the Facial Action Coding System (FACS) and/or perceptual studies to evaluate the intensity of the retargeted avatar facial expression.

\section{Access and Citation}
 The \textit{avatar} is available for distribution to researchers online.  The avatar, code, models and software information can be found at \url{https://github.com/danmcduff/AvatarSim}. The software is released under the MIT license and a responsible AI license\footnote{https://www.licenses.ai/}. This allows the use of the software without restriction, including without limiting rights to use, copy or modify it, with the exception of a set of explicit applications.  If you find this software helpful or use it in research, please cite this paper. Citation information can be found on the GITHUB page.
 
 \section{Conclusion}
 We present an open high-fidelity embodied avatar with bone, action unit, and phoneme controls. We release code and models to illustrate how the avatar can be controlled like a puppet, via expression retargeting or used to create a simple conversational agent using public APIs. We hope that this resource enables researchers to advance the state of the art in high-fidelity embodied virtual agents. Contributions to the code repository are welcomed.

\balance{}
\pagebreak

\bibliographystyle{ACM-Reference-Format}
\bibliography{main}


\begin{thebibliography}{28}


\ifx \showCODEN    \undefined \def \showCODEN     #1{\unskip}     \fi
\ifx \showDOI      \undefined \def \showDOI       #1{#1}\fi
\ifx \showISBNx    \undefined \def \showISBNx     #1{\unskip}     \fi
\ifx \showISBNxiii \undefined \def \showISBNxiii  #1{\unskip}     \fi
\ifx \showISSN     \undefined \def \showISSN      #1{\unskip}     \fi
\ifx \showLCCN     \undefined \def \showLCCN      #1{\unskip}     \fi
\ifx \shownote     \undefined \def \shownote      #1{#1}          \fi
\ifx \showarticletitle \undefined \def \showarticletitle #1{#1}   \fi
\ifx \showURL      \undefined \def \showURL       {\relax}        \fi
\providecommand\bibfield[2]{#2}
\providecommand\bibinfo[2]{#2}
\providecommand\natexlab[1]{#1}
\providecommand\showeprint[2][]{arXiv:#2}

\bibitem[\protect\citeauthoryear{Aneja, Chaudhuri, Colburn, Faigin, Shapiro,
  and Mones}{Aneja et~al\mbox{.}}{2018}]%
        {aneja2018learning}
\bibfield{author}{\bibinfo{person}{Deepali Aneja}, \bibinfo{person}{Bindita
  Chaudhuri}, \bibinfo{person}{Alex Colburn}, \bibinfo{person}{Gary Faigin},
  \bibinfo{person}{Linda Shapiro}, {and} \bibinfo{person}{Barbara Mones}.}
  \bibinfo{year}{2018}\natexlab{}.
\newblock \showarticletitle{Learning to generate 3D stylized character
  expressions from humans}. In \bibinfo{booktitle}{\emph{2018 IEEE Winter
  Conference on Applications of Computer Vision (WACV)}}. IEEE,
  \bibinfo{pages}{160--169}.
\newblock


\bibitem[\protect\citeauthoryear{Aneja, Colburn, Faigin, Shapiro, and
  Mones}{Aneja et~al\mbox{.}}{2016}]%
        {aneja2016modeling}
\bibfield{author}{\bibinfo{person}{Deepali Aneja}, \bibinfo{person}{Alex
  Colburn}, \bibinfo{person}{Gary Faigin}, \bibinfo{person}{Linda Shapiro},
  {and} \bibinfo{person}{Barbara Mones}.} \bibinfo{year}{2016}\natexlab{}.
\newblock \showarticletitle{Modeling stylized character expressions via deep
  learning}. In \bibinfo{booktitle}{\emph{Asian conference on computer
  vision}}. Springer, \bibinfo{pages}{136--153}.
\newblock


\bibitem[\protect\citeauthoryear{Cassell}{Cassell}{2001}]%
        {cassell2001embodied}
\bibfield{author}{\bibinfo{person}{Justine Cassell}.}
  \bibinfo{year}{2001}\natexlab{}.
\newblock \showarticletitle{Embodied conversational agents: representation and
  intelligence in user interfaces}.
\newblock \bibinfo{journal}{\emph{AI magazine}} \bibinfo{volume}{22},
  \bibinfo{number}{4} (\bibinfo{year}{2001}), \bibinfo{pages}{67--67}.
\newblock


\bibitem[\protect\citeauthoryear{Cassell, Sullivan, Churchill, and
  Prevost}{Cassell et~al\mbox{.}}{2000}]%
        {cassell2000embodied}
\bibfield{author}{\bibinfo{person}{Justine Cassell}, \bibinfo{person}{Joseph
  Sullivan}, \bibinfo{person}{Elizabeth Churchill}, {and}
  \bibinfo{person}{Scott Prevost}.} \bibinfo{year}{2000}\natexlab{}.
\newblock \bibinfo{booktitle}{\emph{Embodied conversational agents}}.
\newblock \bibinfo{publisher}{MIT press}.
\newblock


\bibitem[\protect\citeauthoryear{Ekman}{Ekman}{1972}]%
        {Ekman1972}
\bibfield{author}{\bibinfo{person}{Paul Ekman}.}
  \bibinfo{year}{1972}\natexlab{}.
\newblock \showarticletitle{{Universals and cultural differences in facial
  expressions of emotion}}.
\newblock In \bibinfo{booktitle}{\emph{Nebraska Symposium on Motivation}},
  \bibfield{editor}{\bibinfo{person}{J~Cole}} (Ed.).
  \bibinfo{publisher}{University of Nebraska Press},
  \bibinfo{address}{Lincoln}, \bibinfo{pages}{207--282}.
\newblock


\bibitem[\protect\citeauthoryear{Fabian Benitez-Quiroz, Srinivasan, and
  Martinez}{Fabian Benitez-Quiroz et~al\mbox{.}}{2016}]%
        {fabian2016emotionet}
\bibfield{author}{\bibinfo{person}{C Fabian Benitez-Quiroz},
  \bibinfo{person}{Ramprakash Srinivasan}, {and} \bibinfo{person}{Aleix~M
  Martinez}.} \bibinfo{year}{2016}\natexlab{}.
\newblock \showarticletitle{Emotionet: An accurate, real-time algorithm for the
  automatic annotation of a million facial expressions in the wild}. In
  \bibinfo{booktitle}{\emph{Proceedings of the IEEE Conference on Computer
  Vision and Pattern Recognition}}. \bibinfo{pages}{5562--5570}.
\newblock


\bibitem[\protect\citeauthoryear{Friesen, Ekman, et~al\mbox{.}}{Friesen
  et~al\mbox{.}}{1983}]%
        {friesen1983emfacs}
\bibfield{author}{\bibinfo{person}{Wallace~V Friesen}, \bibinfo{person}{Paul
  Ekman}, {et~al\mbox{.}}} \bibinfo{year}{1983}\natexlab{}.
\newblock \showarticletitle{EMFACS-7: Emotional facial action coding system}.
\newblock \bibinfo{journal}{\emph{Unpublished manuscript, University of
  California at San Francisco}} \bibinfo{volume}{2}, \bibinfo{number}{36}
  (\bibinfo{year}{1983}), \bibinfo{pages}{1}.
\newblock


\bibitem[\protect\citeauthoryear{Games}{Games}{2007}]%
        {games2007unreal}
\bibfield{author}{\bibinfo{person}{Epic Games}.}
  \bibinfo{year}{2007}\natexlab{}.
\newblock \showarticletitle{Unreal engine}.
\newblock \bibinfo{journal}{\emph{Online: https://www. unrealengine. com}}
  (\bibinfo{year}{2007}).
\newblock


\bibitem[\protect\citeauthoryear{Gilbert, Demarchi, and Urdapilleta}{Gilbert
  et~al\mbox{.}}{2018}]%
        {gilbert2018facshuman}
\bibfield{author}{\bibinfo{person}{Micha{\"e}l Gilbert},
  \bibinfo{person}{Samuel Demarchi}, {and} \bibinfo{person}{Isabel
  Urdapilleta}.} \bibinfo{year}{2018}\natexlab{}.
\newblock \showarticletitle{FACSHuman a Software to Create Experimental
  Material by Modeling 3D Facial Expression}. In
  \bibinfo{booktitle}{\emph{Proceedings of the 18th International Conference on
  Intelligent Virtual Agents}}. ACM, \bibinfo{pages}{333--334}.
\newblock


\bibitem[\protect\citeauthoryear{He, Zhang, Ren, and Sun}{He
  et~al\mbox{.}}{2016}]%
        {he2016deep}
\bibfield{author}{\bibinfo{person}{Kaiming He}, \bibinfo{person}{Xiangyu
  Zhang}, \bibinfo{person}{Shaoqing Ren}, {and} \bibinfo{person}{Jian Sun}.}
  \bibinfo{year}{2016}\natexlab{}.
\newblock \showarticletitle{Deep residual learning for image recognition}. In
  \bibinfo{booktitle}{\emph{Proceedings of the IEEE conference on computer
  vision and pattern recognition}}. \bibinfo{pages}{770--778}.
\newblock


\bibitem[\protect\citeauthoryear{Hoegen, Aneja, McDuff, and Czerwinski}{Hoegen
  et~al\mbox{.}}{2019}]%
        {hoegen2019end}
\bibfield{author}{\bibinfo{person}{Rens Hoegen}, \bibinfo{person}{Deepali
  Aneja}, \bibinfo{person}{Daniel McDuff}, {and} \bibinfo{person}{Mary
  Czerwinski}.} \bibinfo{year}{2019}\natexlab{}.
\newblock \showarticletitle{An End-to-End Conversational Style Matching Agent}.
\newblock \bibinfo{journal}{\emph{arXiv preprint arXiv:1904.02760}}
  (\bibinfo{year}{2019}).
\newblock


\bibitem[\protect\citeauthoryear{Huber, McDuff, Brockett, Galley, and
  Dolan}{Huber et~al\mbox{.}}{2018}]%
        {huber2018emotional}
\bibfield{author}{\bibinfo{person}{Bernd Huber}, \bibinfo{person}{Daniel
  McDuff}, \bibinfo{person}{Chris Brockett}, \bibinfo{person}{Michel Galley},
  {and} \bibinfo{person}{Bill Dolan}.} \bibinfo{year}{2018}\natexlab{}.
\newblock \showarticletitle{Emotional dialogue generation using image-grounded
  language models}. In \bibinfo{booktitle}{\emph{Proceedings of the 2018 CHI
  Conference on Human Factors in Computing Systems}}. ACM,
  \bibinfo{pages}{277}.
\newblock


\bibitem[\protect\citeauthoryear{Huggins-Daines, Kumar, Chan, Black,
  Ravishankar, and Rudnicky}{Huggins-Daines et~al\mbox{.}}{2006}]%
        {huggins2006pocketsphinx}
\bibfield{author}{\bibinfo{person}{David Huggins-Daines},
  \bibinfo{person}{Mohit Kumar}, \bibinfo{person}{Arthur Chan},
  \bibinfo{person}{Alan~W Black}, \bibinfo{person}{Mosur Ravishankar}, {and}
  \bibinfo{person}{Alexander~I Rudnicky}.} \bibinfo{year}{2006}\natexlab{}.
\newblock \showarticletitle{Pocketsphinx: A free, real-time continuous speech
  recognition system for hand-held devices}. In \bibinfo{booktitle}{\emph{2006
  IEEE International Conference on Acoustics Speech and Signal Processing
  Proceedings}}, Vol.~\bibinfo{volume}{1}. IEEE, \bibinfo{pages}{I--I}.
\newblock


\bibitem[\protect\citeauthoryear{Interuniversitario}{Interuniversitario}{[n.
  d.]}]%
        {interuniversitariolucia}
\bibfield{author}{\bibinfo{person}{CINECA-Consorzio Interuniversitario}.}
  \bibinfo{year}{[n. d.]}\natexlab{}.
\newblock \showarticletitle{LUCIA: An open source 3D expressive avatar for
  multimodal hmi}.
\newblock  (\bibinfo{year}{[n. d.]}).
\newblock


\bibitem[\protect\citeauthoryear{King}{King}{2009}]%
        {dlib09}
\bibfield{author}{\bibinfo{person}{Davis~E. King}.}
  \bibinfo{year}{2009}\natexlab{}.
\newblock \showarticletitle{Dlib-ml: A Machine Learning Toolkit}.
\newblock \bibinfo{journal}{\emph{Journal of Machine Learning Research}}
  \bibinfo{volume}{10} (\bibinfo{year}{2009}), \bibinfo{pages}{1755--1758}.
\newblock


\bibitem[\protect\citeauthoryear{Li, Mehta, Aneja, Foster, Ventola, Shic, and
  Shapiro}{Li et~al\mbox{.}}{2019}]%
        {auModel}
\bibfield{author}{\bibinfo{person}{Beibin Li}, \bibinfo{person}{Sachin Mehta},
  \bibinfo{person}{Deepali Aneja}, \bibinfo{person}{Claire Foster},
  \bibinfo{person}{Pamela Ventola}, \bibinfo{person}{Frederick Shic}, {and}
  \bibinfo{person}{Linda Shapiro}.} \bibinfo{year}{2019}\natexlab{}.
\newblock \bibinfo{title}{A Facial Affect Analysis System for Autism Spectrum
  Disorder}.
\newblock
\newblock
\showeprint{arXiv:1904.03616}


\bibitem[\protect\citeauthoryear{Ma, Mcduff, and Song}{Ma
  et~al\mbox{.}}{2018}]%
        {ma2018generative}
\bibfield{author}{\bibinfo{person}{Shuang Ma}, \bibinfo{person}{Daniel Mcduff},
  {and} \bibinfo{person}{Yale Song}.} \bibinfo{year}{2018}\natexlab{}.
\newblock \showarticletitle{A generative adversarial network for style modeling
  in a text-to-speech system}.
\newblock  (\bibinfo{year}{2018}).
\newblock


\bibitem[\protect\citeauthoryear{McDonnell, Breidt, and B{\"u}lthoff}{McDonnell
  et~al\mbox{.}}{2012}]%
        {mcdonnell2012render}
\bibfield{author}{\bibinfo{person}{Rachel McDonnell}, \bibinfo{person}{Martin
  Breidt}, {and} \bibinfo{person}{Heinrich~H B{\"u}lthoff}.}
  \bibinfo{year}{2012}\natexlab{}.
\newblock \showarticletitle{Render me real?: investigating the effect of render
  style on the perception of animated virtual humans}.
\newblock \bibinfo{journal}{\emph{ACM Transactions on Graphics (TOG)}}
  \bibinfo{volume}{31}, \bibinfo{number}{4} (\bibinfo{year}{2012}),
  \bibinfo{pages}{91}.
\newblock


\bibitem[\protect\citeauthoryear{McDuff, Mahmoud, Mavadati, Amr, Turcot, and
  Kaliouby}{McDuff et~al\mbox{.}}{2016}]%
        {mcduff2016affdex}
\bibfield{author}{\bibinfo{person}{Daniel McDuff}, \bibinfo{person}{Abdelrahman
  Mahmoud}, \bibinfo{person}{Mohammad Mavadati}, \bibinfo{person}{May Amr},
  \bibinfo{person}{Jay Turcot}, {and} \bibinfo{person}{Rana~el Kaliouby}.}
  \bibinfo{year}{2016}\natexlab{}.
\newblock \showarticletitle{AFFDEX SDK: a cross-platform real-time multi-face
  expression recognition toolkit}. In \bibinfo{booktitle}{\emph{Proceedings of
  the 2016 CHI conference extended abstracts on human factors in computing
  systems}}. ACM, \bibinfo{pages}{3723--3726}.
\newblock


\bibitem[\protect\citeauthoryear{Paszke, Gross, Chintala, Chanan, Yang, DeVito,
  Lin, Desmaison, Antiga, and Lerer}{Paszke et~al\mbox{.}}{2017}]%
        {paszke2017automatic}
\bibfield{author}{\bibinfo{person}{Adam Paszke}, \bibinfo{person}{Sam Gross},
  \bibinfo{person}{Soumith Chintala}, \bibinfo{person}{Gregory Chanan},
  \bibinfo{person}{Edward Yang}, \bibinfo{person}{Zachary DeVito},
  \bibinfo{person}{Zeming Lin}, \bibinfo{person}{Alban Desmaison},
  \bibinfo{person}{Luca Antiga}, {and} \bibinfo{person}{Adam Lerer}.}
  \bibinfo{year}{2017}\natexlab{}.
\newblock \showarticletitle{Automatic differentiation in PyTorch}.
\newblock  (\bibinfo{year}{2017}).
\newblock


\bibitem[\protect\citeauthoryear{Pelachaud and Poggi}{Pelachaud and
  Poggi}{2002}]%
        {pelachaud2002subtleties}
\bibfield{author}{\bibinfo{person}{Catherine Pelachaud} {and}
  \bibinfo{person}{Isabella Poggi}.} \bibinfo{year}{2002}\natexlab{}.
\newblock \showarticletitle{Subtleties of facial expressions in embodied
  agents}.
\newblock \bibinfo{journal}{\emph{The Journal of Visualization and Computer
  Animation}} \bibinfo{volume}{13}, \bibinfo{number}{5} (\bibinfo{year}{2002}),
  \bibinfo{pages}{301--312}.
\newblock


\bibitem[\protect\citeauthoryear{Poggi, Pelachaud, de~Rosis, Carofiglio, and
  De~Carolis}{Poggi et~al\mbox{.}}{2005}]%
        {poggi2005greta}
\bibfield{author}{\bibinfo{person}{Isabella Poggi}, \bibinfo{person}{Catherine
  Pelachaud}, \bibinfo{person}{Fiorella de Rosis}, \bibinfo{person}{Valeria
  Carofiglio}, {and} \bibinfo{person}{Berardina De~Carolis}.}
  \bibinfo{year}{2005}\natexlab{}.
\newblock \showarticletitle{Greta. a believable embodied conversational agent}.
\newblock In \bibinfo{booktitle}{\emph{Multimodal intelligent information
  presentation}}. \bibinfo{publisher}{Springer}, \bibinfo{pages}{3--25}.
\newblock


\bibitem[\protect\citeauthoryear{Santos-P{\'e}rez, Gonz{\'a}lez-Parada, and
  Cano-Garc{\'\i}a}{Santos-P{\'e}rez et~al\mbox{.}}{2011}]%
        {santos2011avatar}
\bibfield{author}{\bibinfo{person}{Marcos Santos-P{\'e}rez},
  \bibinfo{person}{Eva Gonz{\'a}lez-Parada}, {and}
  \bibinfo{person}{Jos{\'e}~Manuel Cano-Garc{\'\i}a}.}
  \bibinfo{year}{2011}\natexlab{}.
\newblock \showarticletitle{AVATAR: An Open Source Architecture for Embodied
  Conversational Agents in Smart Environments}. In
  \bibinfo{booktitle}{\emph{International Workshop on Ambient Assisted
  Living}}. Springer, \bibinfo{pages}{109--115}.
\newblock


\bibitem[\protect\citeauthoryear{Serban, Sordoni, Bengio, Courville, and
  Pineau}{Serban et~al\mbox{.}}{2016}]%
        {serban2016building}
\bibfield{author}{\bibinfo{person}{Iulian~V Serban},
  \bibinfo{person}{Alessandro Sordoni}, \bibinfo{person}{Yoshua Bengio},
  \bibinfo{person}{Aaron Courville}, {and} \bibinfo{person}{Joelle Pineau}.}
  \bibinfo{year}{2016}\natexlab{}.
\newblock \showarticletitle{Building end-to-end dialogue systems using
  generative hierarchical neural network models}. In
  \bibinfo{booktitle}{\emph{Thirtieth AAAI Conference on Artificial
  Intelligence}}.
\newblock


\bibitem[\protect\citeauthoryear{Shah, Dey, Lovett, and Kapoor}{Shah
  et~al\mbox{.}}{2018}]%
        {shah2018airsim}
\bibfield{author}{\bibinfo{person}{Shital Shah}, \bibinfo{person}{Debadeepta
  Dey}, \bibinfo{person}{Chris Lovett}, {and} \bibinfo{person}{Ashish Kapoor}.}
  \bibinfo{year}{2018}\natexlab{}.
\newblock \showarticletitle{Airsim: High-fidelity visual and physical
  simulation for autonomous vehicles}. In \bibinfo{booktitle}{\emph{Field and
  service robotics}}. Springer, \bibinfo{pages}{621--635}.
\newblock


\bibitem[\protect\citeauthoryear{Skerry-Ryan, Battenberg, Xiao, Wang, Stanton,
  Shor, Weiss, Clark, and Saurous}{Skerry-Ryan et~al\mbox{.}}{2018}]%
        {skerry2018towards}
\bibfield{author}{\bibinfo{person}{RJ Skerry-Ryan}, \bibinfo{person}{Eric
  Battenberg}, \bibinfo{person}{Ying Xiao}, \bibinfo{person}{Yuxuan Wang},
  \bibinfo{person}{Daisy Stanton}, \bibinfo{person}{Joel Shor},
  \bibinfo{person}{Ron~J Weiss}, \bibinfo{person}{Rob Clark}, {and}
  \bibinfo{person}{Rif~A Saurous}.} \bibinfo{year}{2018}\natexlab{}.
\newblock \showarticletitle{Towards end-to-end prosody transfer for expressive
  speech synthesis with tacotron}.
\newblock \bibinfo{journal}{\emph{arXiv preprint arXiv:1803.09047}}
  (\bibinfo{year}{2018}).
\newblock


\bibitem[\protect\citeauthoryear{van~der Struijk, Huang, Mirzaei, and
  Nishida}{van~der Struijk et~al\mbox{.}}{2018}]%
        {van2018facsvatar}
\bibfield{author}{\bibinfo{person}{Stef van~der Struijk},
  \bibinfo{person}{Hung-Hsuan Huang}, \bibinfo{person}{Maryam~Sadat Mirzaei},
  {and} \bibinfo{person}{Toyoaki Nishida}.} \bibinfo{year}{2018}\natexlab{}.
\newblock \showarticletitle{FACSvatar: An Open Source Modular Framework for
  Real-Time FACS based Facial Animation}. In
  \bibinfo{booktitle}{\emph{Proceedings of the 18th International Conference on
  Intelligent Virtual Agents}}. ACM, \bibinfo{pages}{159--164}.
\newblock


\bibitem[\protect\citeauthoryear{Xiong, Wu, Alleva, Droppo, Huang, and
  Stolcke}{Xiong et~al\mbox{.}}{2018}]%
        {xiong2018microsoft}
\bibfield{author}{\bibinfo{person}{Wayne Xiong}, \bibinfo{person}{Lingfeng Wu},
  \bibinfo{person}{Fil Alleva}, \bibinfo{person}{Jasha Droppo},
  \bibinfo{person}{Xuedong Huang}, {and} \bibinfo{person}{Andreas Stolcke}.}
  \bibinfo{year}{2018}\natexlab{}.
\newblock \showarticletitle{The Microsoft 2017 conversational speech
  recognition system}. In \bibinfo{booktitle}{\emph{2018 IEEE International
  Conference on Acoustics, Speech and Signal Processing (ICASSP)}}. IEEE,
  \bibinfo{pages}{5934--5938}.
\newblock


\end{thebibliography}

\end{document}